\documentclass[aip,rsi,graphicx,reprint]{revtex4-1}
\usepackage{amssymb}
\usepackage{graphicx}% Include figure files
\usepackage{dcolumn}% Align table columns on decimal point
\usepackage{bm}% bold math
\usepackage[utf8]{inputenc}
\usepackage[T1]{fontenc}
\usepackage{mathptmx}
\usepackage{etoolbox}
\usepackage{multirow}

\begin{document}
% Use the \preprint command to place your local institutional report number 
% on the title page in preprint mode.
% Multiple \preprint commands are allowed.
%\preprint{}

\title{High-Q Superconducting Lumped-Element Resonators for Low-Mass Axion Searches} %Title of paper

% repeat the \author .. \affiliation  etc. as needed
% \email, \thanks, \homepage, \altaffiliation all apply to the current author.
% Explanatory text should go in the []'s, 
% actual e-mail address or url should go in the {}'s for \email and \homepage.
% Please use the appropriate macro for the type of information

% \affiliation command applies to all authors since the last \affiliation command. 
% The \affiliation command should follow the other information.
\author{Roman Kolevatov}
\email[Corresponding author: ]{rkolevatov@princeton.edu}
%\homepage[]{Your web page}
%\thanks{}
%\altaffiliation{}
\affiliation{Department of Physics, Princeton University, Princeton, New Jersey 08544, USA}

\author{Saptarshi Chaudhuri}
%\homepage[]{Your web page}
%\thanks{}
%\altaffiliation{}
\affiliation{Department of Physics, Princeton University, Princeton, New Jersey 08544, USA}

\author{Lyman Page}
%\homepage[]{Your web page}
%\thanks{}
%\altaffiliation{}
\affiliation{Department of Physics, Princeton University, Princeton, New Jersey 08544, USA}

% Collaboration name, if desired (requires use of superscriptaddress option in \documentclass). 
% \noaffiliation is required (may also be used with the \author command).
%\collaboration{}
%\noaffiliation

\date{\today}

\begin{abstract}
Low-frequency superconducting lumped-element resonators have recently attracted significant attention in the context of axion dark matter searches. Here we present the design and implementation of a fixed-frequency superconducting resonator operating near \(250~\mathrm{kHz}\), possessing an inductor volume of $\sim 1$ liter and achieving an unloaded quality factor \(Q \approx 2.1\times10^{6}\). This resonator represents a significant improvement over the state of the art and informs the design of searches for low-mass axions.
\end{abstract}

% insert suggested PACS numbers in braces on next line

\maketitle %\maketitle must follow title, authors, abstract and \pacs

% Body of paper goes here. Use proper sectioning commands. 
% References should be done using the \cite, \ref, and \label commands
\section{Introduction}
\label{sec:intro}
Resonators are ubiquitous devices in experimental physics and electrical engineering.\cite{Braginsky1985SmallDissipation} They store energy and respond strongly at the resonant frequency, so signals near that frequency are enhanced. The two main figures of merit are the resonance frequency and the quality factor \(Q\), which sets the bandwidth and quantifies loss.\cite{Pozar2012}

Before presenting the \(250~\mathrm{kHz}\) resonator--the main result of this work--we briefly review the state of the art at higher frequencies and motivate the need for high-\(Q\) resonators in low-mass (frequency) axion searches.

Substantial progress in developing high-quality (high-\(Q\)) resonators has been achieved at gigahertz frequencies. For example, superconducting radio-frequency (SRF) cavities for high-energy accelerators\cite{Padamsee2008_RFSuperconductivity} routinely attain \(Q\approx10^{10}\).\cite{Padamsee2014, Romanenko_Grassellino2014} SRF technology is well developed and relies on specific processes such as surface electropolishing,\cite{Kneisel2006} high-pressure water rinsing and clean-room assembly,\cite{DESY2008,FNAL2018} heat treatments,\cite{Ciovati2004,Romanenko2013} and nitrogen doping.\cite{Grassellino2013,Gonnella2016} Another rapidly developing area requiring high-\(Q\) gigahertz resonators is quantum information science. In circuit quantum electrodynamics (cQED), 3D superconducting cavities operated at the single-photon level have demonstrated \(Q\approx10^{9}\).\cite{Reagor2013,Brecht2016}

In recent years, high-\(Q\) resonators have become increasingly important in experimental cosmology due to the interest in axion dark matter.\cite{adams2023axiondarkmatter,PhysRevD.110.030001} Axions can be treated as a classical wave with a Compton frequency \(\nu_a=m_a c^2/h\). Searches in the mass range \(m_a \sim 1~\mu\mathrm{eV}\text{--}0.1~\mathrm{meV}\) naturally target resonant microwave cavities (resonant frequencies from hundreds of MHz to tens of GHz). The resonant microwave-cavity haloscope was proposed by Pierre Sikivie to probe the axion-photon coupling and remains the leading technique in this band.\cite{Sikivie1983} Experiments such as HAYSTAC,\cite{Zhong2018,Jewell2023,Bai2025} ADMX,\cite{Du2018,Braine2020,Bartram2021,Goodman2025} and CAPP,\cite{PhysRevLett.124.101802,PhysRevLett.130.071002,PhysRevX.14.031023} use cylindrical cavities to resonantly enhance the weak axion-conversion signal. High \(Q\) for these cavities is essential: the expected signal power--and thus the scan rate--scales linearly with \(Q\).\cite{Sikivie2021}

In contrast to the mature technologies of gigahertz SRF and 3D-qubit cavities, and the substantial knowledge base for cylindrical microwave cavities, low-frequency \emph{lumped-element resonators} in the \(\mathrm{kHz}\text{-}\mathrm{MHz} \) frequency range remain comparatively underdeveloped. At $\sim$1 kHz, quality factors on the order of \(10^{6}\) have been demonstrated--for example, \(Q\approx 1.8\times10^{6}\) in the \(145\text{--}175~\mathrm{Hz}\) band,\cite{Bonaldi1998} and \(Q\approx 1.6\times10^{6}\) in the \(250\text{--}1500~\mathrm{Hz}\) band,\cite{Falferi1994} using a superconducting LC resonator as a transducer to read out a signal from a Weber bar in gravitational wave searches. At hundreds of kHz, \(Q\) typically degrades as dielectric losses become more significant, often scaling approximately with the square of frequency.\cite{Falferi1994} Consistent with this trend, for resonators used in SQUID-based low-field MRI,\cite{Tesche1977,Clarke1979,Clarke2007} the quality factor reached \(Q \approx 5.55\times10^{4}\) at \(425~\mathrm{kHz}\).\cite{Myers2007}. Similarly, \(Q \approx 4\times10^{4}\) was achieved at \(492~\mathrm{kHz}\) in a search for hidden photon dark matter \cite{phipps2020exclusion}. Superconducting resonators used in a multi-Penning-trap system have reached \(Q\gtrsim 8\times10^{4}\) in the \(520\text{--}710~\mathrm{kHz}\),\cite{Volksen2022} and an unloaded \(Q\approx 5\times10^{5}\) has been achieved in the \(550\text{--}800~\mathrm{kHz}\) range.\cite{Nagahama2016} Notably, while the latter resonator is of comparable frequency to the one presented here, its volume is an order of magnitude smaller--an important distinction for low-mass axion searches. Larger resonators readily couple more energy from larger magnets, increasing the scan rate\cite{Brouwer2022}. 

Besides extensive microwave-band searches for the axion-photon coupling, lower-mass QCD axions (\(\sim\!\mathrm{peV}\)-\(100~\mathrm{neV}\)) are motivated by the axiverse,\cite{Svrcek2006,Arvanitaki2010} low-scale inflation,\cite{Takahashi2018,Graham2018} and early-universe entropy production.\cite{Steinhardt1983} In this low-mass regime, lumped-element resonators--though less developed--are essential. Simply scaling cylindrical cavities is impractical. The axion Compton wavelength is \(\lambda_a=h/(m_a c)\), so for \(m_a=1~\mathrm{neV}\) one has \(\lambda_a\approx 1.2~\mathrm{km}\). For the commonly used TM\(_{010}\) mode, the cavity radius scales as \(a\approx 0.38\,\lambda\), implying kilometer-scale diameters; for lighter axions, the required dimensions grow \(\propto m_a^{-1}\), rendering cavity haloscopes infeasible and motivating lumped-element (LC) resonators. 

%Examples of low-mass axion searches include CASPEr and DMRadio. CASPer is an NMR-based axion search program,\cite{Budker2014,Garcon2017} which spans a broad frequency range from a few hertz to hundreds of megahertz.

%Several searches for the axion-photon coupling using LC resonators have been proposed and/or operated, including

Examples of searches for the axion-photon coupling using LC resonators include ADMX SLIC\cite{crisosto2020admx}, BASE\cite{devlin2021constraints}, WISPLC\cite{zhang2022search}, and DMRadio\cite{brouwer2022projected,Brouwer2022}. Re-entrant cavities\cite{PhysRevD.94.042001,PhysRevD.109.042004} provide an alternative 3D lumped-element implementation, with spatially separated electric and magnetic fields. (See also \cite{aybas2021quantum}, which describes the use of LC resonators to probe the axion-gluon coupling using NMR-based techniques, e.g., the CASPEr experiment.) As shown in \cite{Brouwer2022}, searching over \(0.4\text{--}120~\mathrm{neV}\) (\(100~\mathrm{kHz}\text{--}30~\mathrm{MHz}\)) with sensitivity to GUT-scale QCD axions requires, among other advances, a tunable lumped-element resonator with quality factor  \(Q=2\times 10^{6}-2\times10^{7}\).\cite{Brouwer2022} Existing resonator technology reported in the literature falls short of this requirement.

%DMRadio is a series of low-mass axion searches that probe the axion-photon coupling with high-\(Q\) lumped-element resonators. The DMRadio-50L experiment plans to employ a high-\(Q\) tunable LC resonator and targets \(20~\mathrm{peV}\text{--}20~\mathrm{neV}\) (rest-mass frequencies \(5~\mathrm{kHz}\text{--}5~\mathrm{MHz}\)).\cite{Rapidis2022} Building on this concept, the proposed DMRadio-GUT aims to search over \(0.4\text{--}120~\mathrm{neV}\) (\(100~\mathrm{kHz}\text{--}30~\mathrm{MHz}\)) with sensitivity to GUT-scale QCD axions; achieving this sensitivity requires, among other advances, a tunable lumped-element resonator with quality factor  \(Q=2\times 10^{6}-2\times10^{7}\).\cite{Brouwer2022} Existing resonator technology reported in the literature falls short of this requirement.

In this paper, we demonstrate a superconducting lumped-element resonator with fixed resonance frequency \(f_{0}\approx 250~\mathrm{kHz}\) and, to our knowledge, an unprecedented unloaded (intrinsic) quality factor \(Q_\mathrm{ul}\approx 2.1\times10^{6}\). This device also serves as a proof-of-concept for forthcoming experiments based on lumped-element resonators and as a practical ``set of recipes'' for future resonator development within low-mass axion searches.

\section{Methods}
\subsection{Resonator Overview \& Assembly}
\begin{table}[t]
\caption{\label{tab:defs}Definitions of the resonator parameters used in this work.}
\begin{ruledtabular}
\begin{tabular}{@{}l p{0.78\linewidth}@{}}
Symbol & Definition \\
\hline
$Q$ &
Loaded quality factor:
\( Q \equiv (\omega_0 L)/R \),
where \(R\) is the total loaded resistance of the resonator circuit coupled to the injection and readout lines [Eq.~(\ref{eq:R_resonator})]. \\[0.6em]

$Q_{\mathrm{ul}}$ &
Unloaded (internal) quality factor:
\( Q_{\mathrm{ul}} \equiv (\omega_0 L)/R_{\mathrm{ul}} \),
where \(R_{\mathrm{ul}}\) is the unloaded (residual) resistance describing internal dissipation in the absence of external loading. \\[0.8em]

$f_{0}$ &
Resonance frequency, defined as the frequency at which the measured power spectral density exhibits a maximum (equivalently, the peak of the Lorentzian response). \\[0.6em]

$\Delta f_{\mathrm{FWHM}}$ &
Resonance linewidth (full width at half maximum), related to the loaded quality factor by
\( \Delta f_{\mathrm{FWHM}} = f_{0}/Q \). \\[0.6em]

$\tau$ &
Ringdown time constant of the resonant mode amplitude, defined by an exponential decay \(V(t)\propto e^{-t/\tau}\). It is related to the loaded quality factor by
\( \tau = Q/(\pi f_{0}) \). \\
\end{tabular}
\end{ruledtabular}
\end{table}

\begin{figure*}
    \centering
    \includegraphics[width=0.5\linewidth]{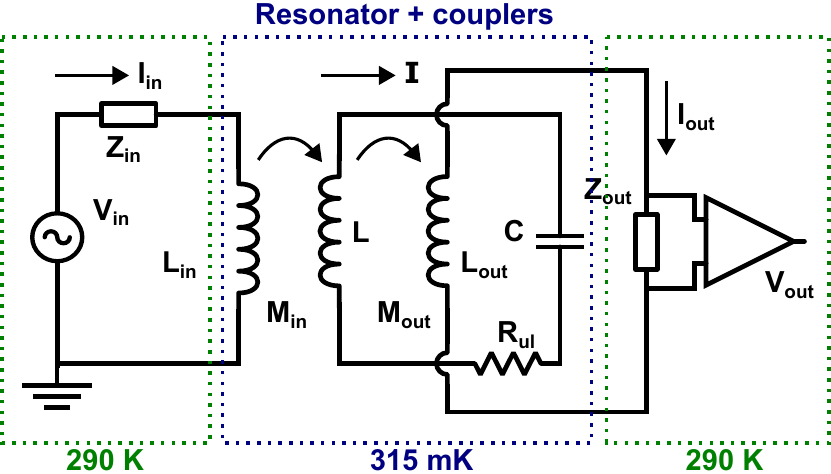}
    \caption{Principal circuit of the resonator with injection and readout lines. A room-temperature function generator applies an input voltage \(V_{\mathrm{in}}\) to the injection line, carrying current \(I_{\mathrm{in}}\). The injection coil is inductively coupled to the resonator coil through mutual inductance \(M_{\mathrm{in}}\), injecting the signal into the resonator at cryogenic temperature (\(\approx 315\)\,mK). The resonator itself consists of an inductor (\(L\)) and capacitor (\(C\)), with an unloaded (residual) resistance \(R_\mathrm{ul}\) that represents internal losses. The output signal is extracted by a readout coil, coupled to the resonator via mutual inductance \(M_{\mathrm{out}}\), producing a current \(I_{\mathrm{out}}\) that travels out of the cryostat and is converted into a voltage \(V_{\mathrm{out}}\) by a room-temperature amplifier. The input and output impedances are \(Z_{\mathrm{in}}\) and \(Z_{\mathrm{out}}\), while the self-inductances of the injection and readout coils are \(L_{\mathrm{in}}\) and \(L_{\mathrm{out}}\).
    }
    \label{fig:resonator_circuit}
\end{figure*}
\begin{figure}
    \centering
    \includegraphics[width=1\linewidth]{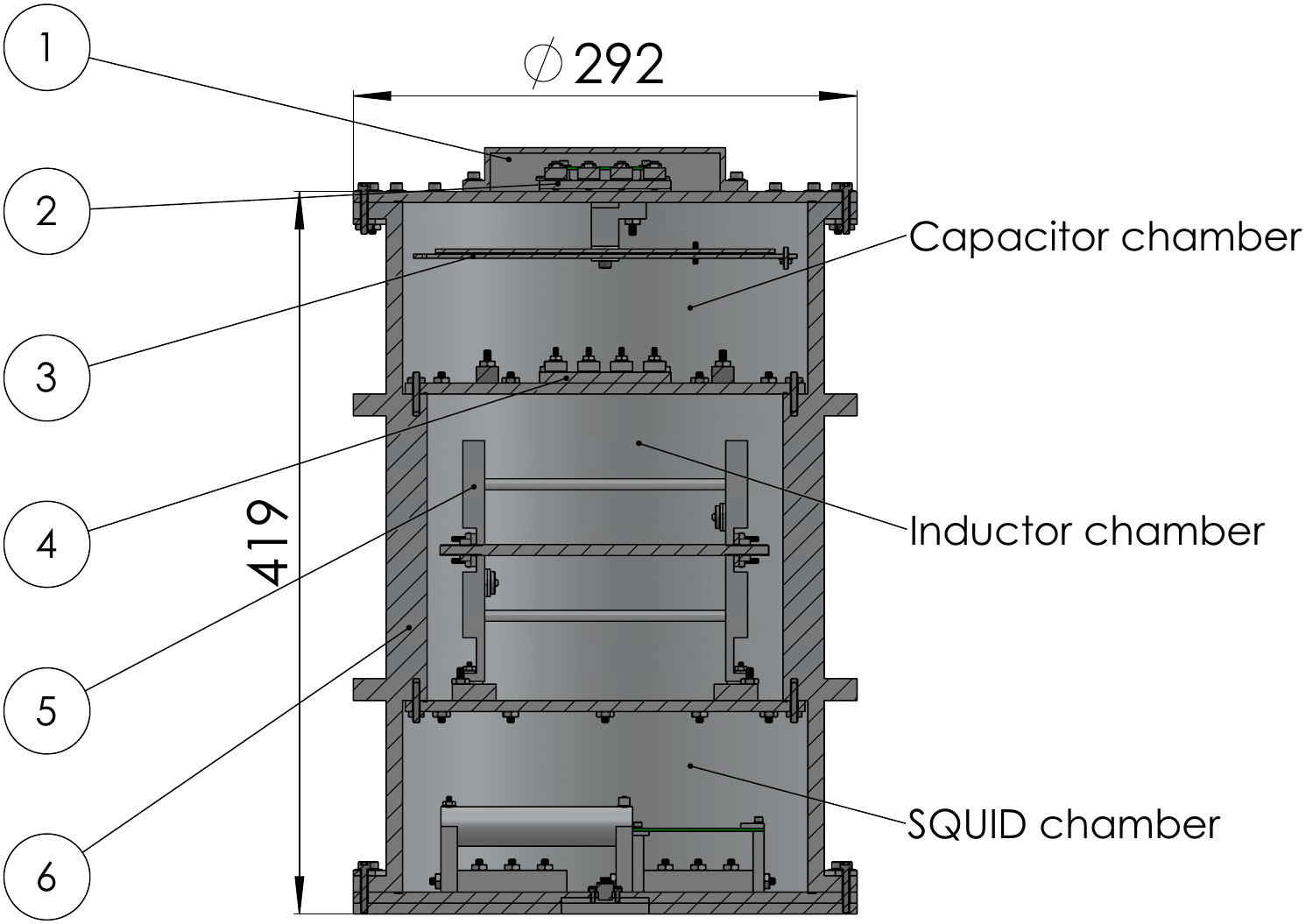}
    \caption{CAD cross section of the resonator apparatus. The dimensions are in millimeters. The top cap (1) encloses screw terminals (2) where the injection and readout lines transition from copper wire (outside the shield) to NbTi (inside). The capacitor (3) is mounted to the ceiling of the capacitor chamber. An intermediate set of screw terminals (4) on a mounting plate inside the capacitor chamber provides a convenient break for the injection and readout lines. The inductor coil (5) is positioned centrally within the inductor chamber and is connected to the capacitor with NbTi leads. A high-purity 5N (99.999\%) aluminum shield (6) surrounds the assembly and defines three chambers: capacitor, inductor, and a SQUID chamber reserved for future integration.}
    \label{fig:CAD_cross_section_resonator}
\end{figure}
\begin{table}[b]
\caption{Measured device parameters.}
\label{tab:R_params}
\begin{ruledtabular}
\begin{tabular}{lcc}
Quantity & Symbol & Value \\
\hline
Self-inductance & \(L\) & \((750 \pm 1)\,\mu\mathrm{H}\) \\
Capacitance     & \(C\) & \((541.9 \pm 0.7)\,\mathrm{pF}\) \\
Resonance frequency & \(f_{0}\) & \( (249{,}656.75 \pm 0.03)\,\mathrm{Hz}\) \\
Injection mutual inductance & \(M_\mathrm{in}\) & \((52 \pm 3)\,\mathrm{nH}\) \\
Readout mutual inductance   & \(M_\mathrm{out}\)   & \((103 \pm 3)\,\mathrm{nH}\) \\
Unloaded resistance         & \(R_\mathrm{ul}\) & \((0.572 \pm 0.006)\,\mathrm{m}\Omega\) \\
Unloaded quality factor     & \(Q_\mathrm{ul}\)           & \((2.06 \pm 0.02)\times 10^{6}\) \\
\end{tabular}
\end{ruledtabular}
\end{table}

The principal circuit diagram of the resonator is shown in Fig.~\ref{fig:resonator_circuit}. The lumped-element resonator comprises a superconducting inductor and a parallel-plate capacitor. For testing the device, the injection and readout lines are inductively coupled to the resonator: the injection line delivers the excitation signal, and the readout line senses the response. For the device studied here, the self-inductance and capacitance are \(L = (750 \pm 1)\,\mu\mathrm{H}\) and \(C = (541.9 \pm 0.7)\,\mathrm{pF}\), corresponding to a resonance frequency \(f_{0} = (249{,}656.75 \pm 0.03)\,\mathrm{Hz}\). 

The total loaded resistance of the resonator circuit, coupled to the injection and readout lines, is
\begin{equation}
\label{eq:R_resonator}
R = R_\mathrm{ul} + \frac{\bigl(\omega M_\mathrm{in}\bigr)^{2}}{Z_\mathrm{in}} + \frac{\bigl(\omega M_\mathrm{out}\bigr)^{2}}{Z_\mathrm{out}},
\end{equation}
where \(\omega = 2\pi f\) and \(f\) is the frequency of the injected signal. Here, \(R_\mathrm{ul}\) denotes the unloaded (residual) resistance of the resonator, representing internal losses in the superconducting inductor, capacitor, and associated wiring. The second term accounts for dissipation due to coupling to the injection line, characterized by its mutual inductance \(M_\mathrm{in}\) and impedance \(Z_\mathrm{in}\); the third term describes analogous losses through the readout line, defined by \(M_\mathrm{out}\) and \(Z_\mathrm{out}\). Equation~(\ref{eq:R_resonator}) assumes \(|Z_\mathrm{in}|\gg \omega L_\mathrm{in}\) and \(|Z_\mathrm{out}|\gg \omega L_\mathrm{out}\), and more generally, that the reactance of the injection and readout lines may be neglected. Then, \(Z_\mathrm{in}\) and \(Z_\mathrm{out}\) may be taken as purely real (i.e., \(Z_\mathrm{in}\approx R_\mathrm{in}\) and \(Z_\mathrm{out}\approx R_\mathrm{out}\)). These conditions are satisfied for the present device. As shown in Table~\ref{tab:R_params}, \(M_\mathrm{in} = (53 \pm 2)\,\mathrm{nH}\) and \(M_\mathrm{out} = (105 \pm 3)\,\mathrm{nH}\). These couplings are kept small to limit the coupling-induced dissipation.

%likewise \(|Z_\mathrm{out}|\gg \omega L_\mathrm{out}\) and \(|Z_\mathrm{out}|\gg 1/(\omega C_\mathrm{out})\); thus the reactive contributions are negligible and 

The CAD drawing showing a cross section of the resonator apparatus is given in Fig.~\ref{fig:CAD_cross_section_resonator}. The resonator has a high-purity 5N (99.999\%) aluminum shield separated into three different chambers with their specific purpose: the top chamber houses the parallel plate capacitor, the middle chamber houses the inductor coil, and the bottom chamber is reserved for the future integration with the SQUID readout (not discussed in this paper).

Material selection was guided by minimizing dissipation while maintaining superconductivity at the operating temperature (\(T \approx 315~\mathrm{mK}\)). All electrical conductors are superconducting. The lead connecting the inductor and capacitor is \(\approx\!5\)-mil (\(127~\mu\mathrm{m}\)) Formvar-insulated niobium-titanium (NbTi) wire \((T_{\mathrm{c}}\approx 9~\mathrm{K})\).\cite{Scanlan2004} Fasteners (studs and nuts) are aluminum alloy (Al~6061), and the bulk resonator parts are commercially pure aluminum (Al~1100). The measured critical temperature \(T_\mathrm{c}\) ranged from \(1.0~\mathrm{K}\) for Al~6061 to \(1.2~\mathrm{K}\) for Al~1100. Selected screws and washers are tantalum \((T_{\mathrm{c}}\approx 4.5~\mathrm{K})\).\cite{Milne1961} Thin-foil indium gaskets \((T_{\mathrm{c}}\approx 3.4~\mathrm{K})\) are used at metal-dielectric interfaces to improve thermal contact.\cite{NIST_SRM_767a_1992}

% Dielectric selection prioritized low microwave loss. All structural dielectrics are single-crystal sapphire (\(\mathrm{Al_2O_3}\)), which has \(\tan\delta \sim 10^{-9}\) at GHz frequencies.\cite{Creedon2011,Read2023}
To minimize microwave loss, all structural dielectrics were implemented as single-crystal sapphire (\(\mathrm{Al_2O_3}\)). At room temperature, polycrystalline alumina (low-purity grade) exhibits \(\tan\delta \sim 10^{-4}\), whereas sapphire reaches \(\tan\delta \sim 10^{-6}\).\cite{Vila1998} The only additional dielectric is the thin Formvar coating on the NbTi lead. Although Formvar exhibits a higher loss tangent (\(\tan\delta \sim 10^{-3}\)),\cite{Ulmer2009} its volume fraction is small, and uninsulated wiring is impractical in the present assembly due to the risk of electrical shorts. Use of uninsulated leads may be explored in future work.

All electrical connections inside the resonator are made with screw terminals, forming superconductor-superconductor (S-S) joints. Soldered joints are avoided because the fine filaments of NbTi are difficult to wet: conventional Sn-Pb solder does not wet NbTi and tends to bead, and flux residues can introduce lossy contaminants that degrade the quality factor \(Q\). At each contact, the Formvar insulation is stripped from the NbTi lead and the conductor is clamped under a tantalum (Ta) washer. The inductor-capacitor connection is made at the capacitor plate's screw terminals using Ta washers and Al~6061 studs and nuts. For assembly convenience, the injection and readout leads terminate at intermediate screw terminals inside the capacitor chamber; these joints also use Ta washers with Al~6061 studs and nuts. Each intermediate terminal is an Al~1100 block electrically isolated from the grounded shield plate by sapphire washers. Tantalum washers with aluminum nuts and studs are preferred at cryogenic temperatures: from \(\sim300\) to \(0~\mathrm{K}\) the integrated linear contraction is \(\Delta L/L \approx 4.15\times 10^{-3}\) for aluminum and \(\approx 1.43\times 10^{-3}\) for tantalum,\cite{NBSMono29} so aluminum contracts more than tantalum during cooldown, increasing the clamping force and tightening the joint.

A previous design employed intermediate screw terminals in the resonator circuit, inserting Al~1100 terminal blocks between the inductor and the capacitor (four connection points). Removal of this intermediate terminal--i.e., direct connection of the inductor to the capacitor--coincided with a substantial improvement in \(Q\), from \(Q\approx2.3\times10^{5}\) to \(Q\approx7.0\times10^{5}\). However, other modifications were implemented in the same cooldown (replacement of alumina with sapphire washers and the addition of indium washers), so the improvement cannot be attributed solely to the removal of the intermediate joint.

Contact preparation is performed immediately prior to assembly. All aluminum parts are sonicated in acetone and then immediately sonicated in \(99\%\) isopropanol (IPA) to minimize residue formation and surface reactions that can occur when acetone is exposed to air. Aluminum surfaces are lightly abraded with silicon-carbide paper (harder than aluminum oxide), progressing from coarse (P200) to fine (P600) grit to break and remove the native oxide \(\mathrm{Al_2O_3}\) layer. The NbTi insulation is removed using an Eraser RT2S magnet-wire stripper (abrasive fiberglass filament). Wire tips are inspected under a microscope before clamping to ensure complete insulation removal and avoid introducing dielectric loss at the joints.

All fasteners were torqued to specification using a calibrated torque wrench. The 8\text{-}32 fasteners were torqued to \(0.54~\mathrm{N\,m}\); the 4\text{-}40 fasteners to \(0.32~\mathrm{N\,m}\) for electrical contacts and \(0.25~\mathrm{N\,m}\) for mechanical contacts; and the three Ta 8\text{-}32 screws clamping the capacitor plates to \(1.13~\mathrm{N\,m}\).
\subsection{Cryogenics}
\begin{figure}
    \centering
    \includegraphics[width=0.75\linewidth]{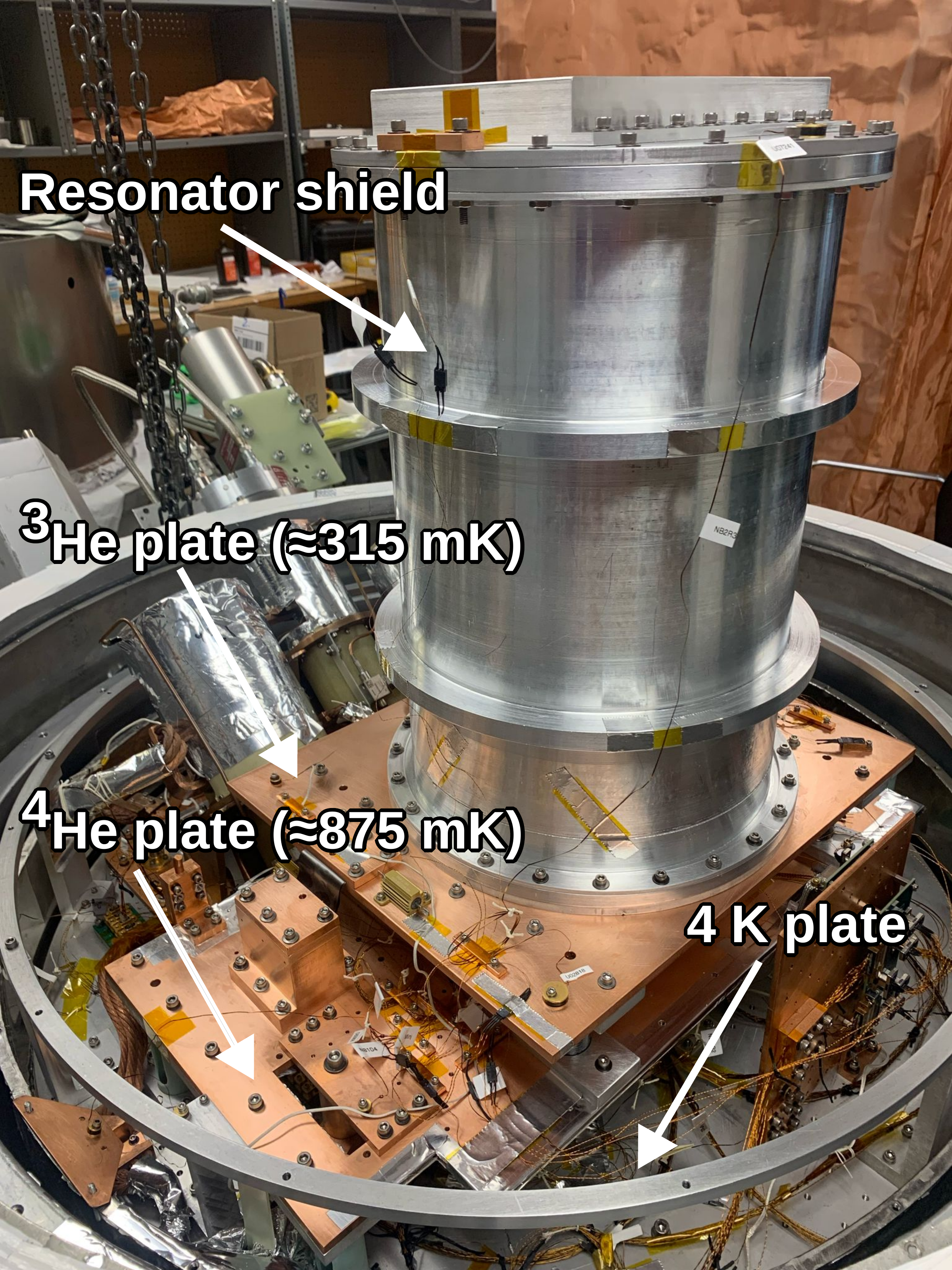}
    \caption{Photograph of the resonator apparatus. The resonator is mounted on a two-level copper table comprising an upper and a lower copper plate. The upper plate is thermally anchored to the bottom of the \(^3\)He pot, while the lower plate is anchored to the bottom of the \(^4\)He pot of the \(^3\)He-\(^4\)He sorption refrigerator. The assembly is mounted to the \(4\,\mathrm{K}\) stage via G-10 support legs; carbon-fiber legs support the upper plate. A passive pyrolytic graphite heat switch thermally links the two plates. The lower copper plate is also coupled to the \(4\,\mathrm{K}\) stage through a second passive pyrolytic-graphite heat switch.%
    }
    \label{fig:photograph_resonator}
\end{figure}
The resonator is cooled using two pulse-tube cryocoolers (PTs): a Cryomech PT410 with a remote motor and a Cryomech PT407,\cite{Wang2002} in combination with a \({}^3\mathrm{He}\)\,--\,\({}^4\mathrm{He}\) sorption refrigerator.\cite{Devlin2004,Lau2006} The PTs provide the primary cooling to a base temperature of \(4\,\mathrm{K}\). Once this temperature is reached, the sorption refrigerator is cycled. The system consists of two \({}^4\mathrm{He}\) pots and one \({}^3\mathrm{He}\) pot.

The resonator apparatus is mounted on a two-level copper platform attached to the \(4~\mathrm{K}\) base stage (Fig.~\ref{fig:photograph_resonator}). The lower copper plate is thermally anchored to the \({}^{4}\mathrm{He}\) pots and reaches \(875~\mathrm{mK}\). The upper plate is supported by carbon-fiber legs--chosen for their low sub-kelvin thermal conductance--and is thermally anchored to the \({}^{3}\mathrm{He}\) pot; this stage reaches \(315~\mathrm{mK}\). The resonator is mounted on this \(315~\mathrm{mK}\) plate, well below the critical temperature \(T_c\) of all superconducting materials used in its construction.

To aid precooling during off-cycle periods, flexible pyrolytic-graphite (PG) sheets are employed as passive heat switches.\cite{Kolevatov2025} The plate anchored to the bottoms of the \({}^4\mathrm{He}\) pots is linked to the \(4\,\mathrm{K}\) base stage with stacks of PG sheets clamped between aluminum plates.
\subsection{Inductor Coil}
\begin{figure}
    \centering
    \includegraphics[width=0.85\linewidth]{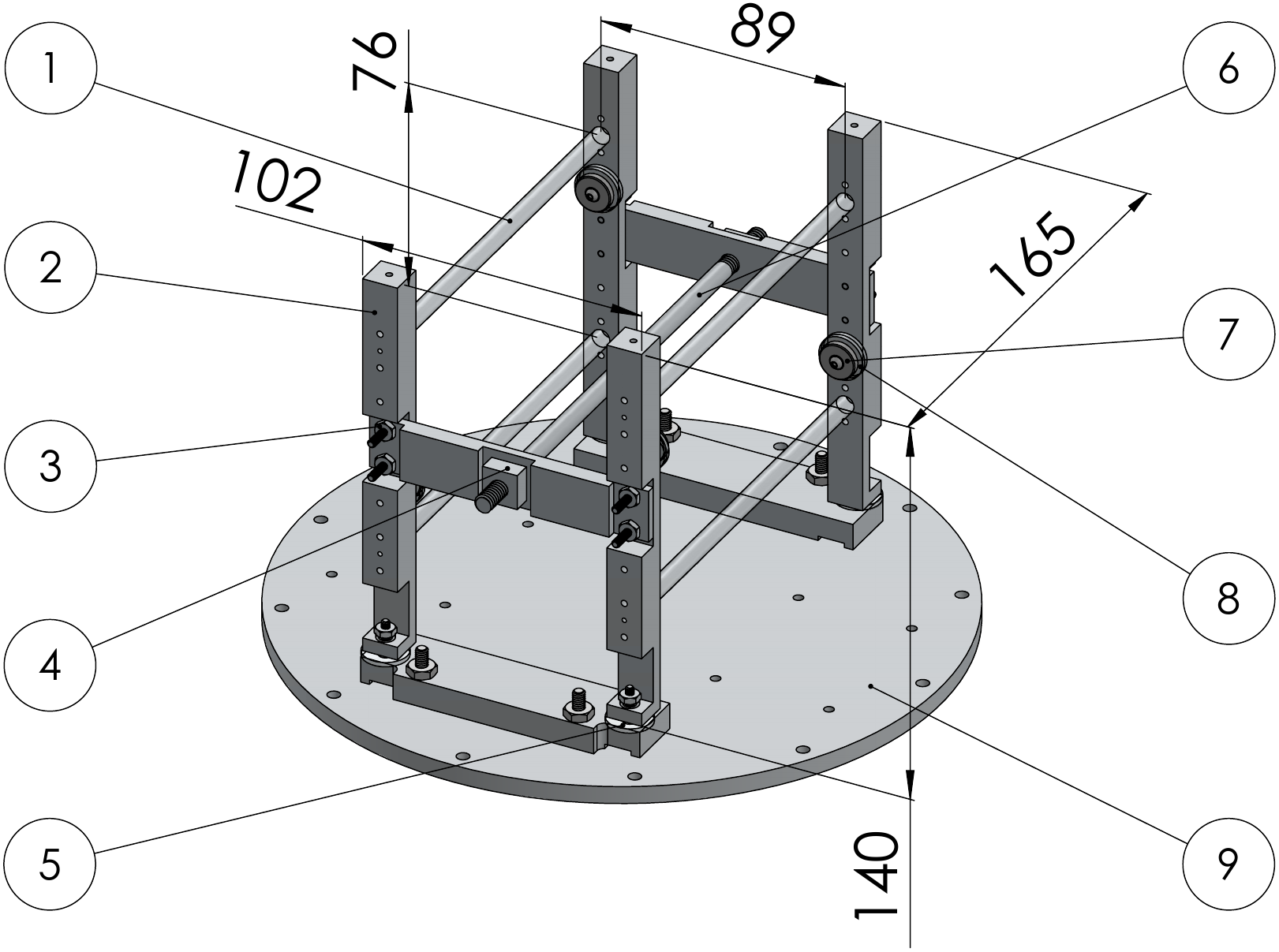}
    \caption{CAD model of the inductor coil assembly. Dimensions are in millimeters. Four sapphire rods (1) join the opposing Al~1100 side frames (2) and serve as winding mandrels for the NbTi wire. The frames are fastened with Al~6061 studs and nuts (3). Al~1100 square nuts (4) secure the ends of a threaded Al~1100 tie rod (6) that clamps the side plates to stiffen the structure. The rod is held in tension and is the only element holding the two ends together. Sapphire washers under the coil (5) provide electrical isolation from the chassis to avoid closed superconducting loops. Each leg carries an Al~6061 flanged-button screw (7) paired with two sapphire washers (8) to anchor the NbTi wire. Two of these screw-washer sets form the injection and readout loops by routing the wire around one sapphire rod, while the remaining two sets terminate the winding. The completed assembly is mounted on an Al~1100 baseplate (9).%
    }
    \label{fig:inductor_CAD}
\end{figure}
\begin{figure}
    \centering
    \includegraphics[width=0.75\linewidth]{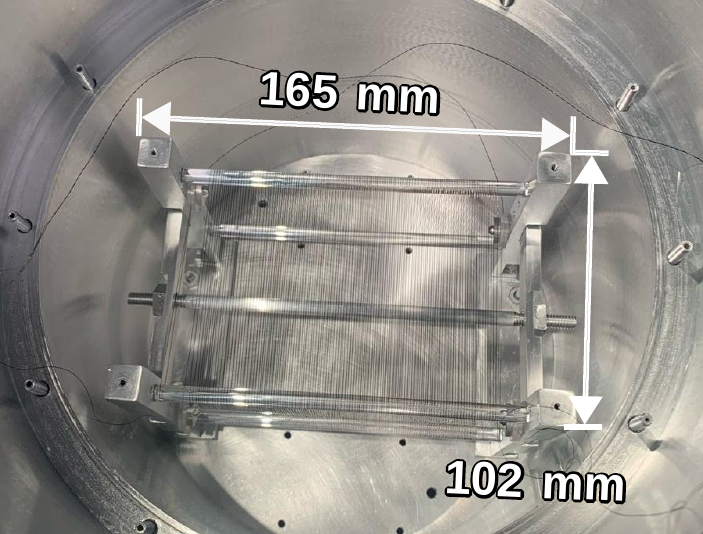}
    \caption{Photograph of the inductor coil mounted inside the high-purity (5N) aluminum resonator shield. When superconducting, the shield expels magnetic field (Meissner effect), screening the flux and thereby lowering the coil’s effective self-inductance.
    }
    \label{fig:inductor_photograph}
\end{figure}
A CAD drawing of the inductor is shown in Fig.~\ref{fig:inductor_CAD}. The frame (Al~1100) consists of two H-shaped end frames tied by four sapphire rods with center-to-center spacings of \(89\,\mathrm{mm}\) (horizontal) and \(76\,\mathrm{mm}\) (vertical), each rod \(6.35\,\mathrm{mm}\) in diameter. The winding is \(127\,\mu\mathrm{m}\) NbTi wire laid on the rods, forming a rectangular solenoid with cross sectional area \(A \approx 7740\,\mathrm{mm^2}\), effective length \(l \approx 121\,\mathrm{mm}\), and \(N \approx 120\) turns. The volume enclosed by the winding is thus roughly 1 liter. A simple estimate \(L \simeq \mu_0 N^{2}A/l\) \((\mu_0 = 4\pi\times10^{-7}\,\mathrm{H\,m^{-1}})\) gives \(L \approx 1160\,\mu\mathrm{H}\), whereas the measured inductance is \(L = (750 \pm 1)\,\mu\mathrm{H}\), consistent with flux screening due to superconducting shielding (see Fig.~\ref{fig:inductor_photograph}).

The Al~1100 frame is electrically isolated from the grounded shield plate by four sapphire washers inserted between the frame legs and the inductor baseplate. This technique prevents the formation of closed superconducting loops that could support persistent screening currents, thereby reducing the coil inductance and distorting the local field.

The coil design underwent several iterations. An initial implementation used an Al~6061 frame with alumina rods. The broader superconducting transition of Al~6061 left residual resistance at \(T\approx 875~\mathrm{mK}\), reducing the unloaded quality factor. Replacing the frame with higher-purity Al~1100--which exhibits a sharper \(T_{\mathrm{c}}\) transition--while retaining the alumina rods increased \(Q\) to \(\approx 7.0\times10^{5}\) at \(875~\mathrm{mK}\) (Fig.~\ref{fig:ringdown}). Substituting sapphire in place of alumina, owing to its lower dielectric loss, raised \(Q\) to \(\approx 1.0\times 10^{6}\) at the same temperature. The largest improvement followed a cryogenic upgrade: adding a \({}^3\mathrm{He}\) pot to the sorption refrigerator enabled operation at \(T \approx 315\,\mathrm{mK}\), yielding the present result \(Q \approx 2.1\times 10^{6}\).

\subsection{Capacitor}
\begin{figure}
    \centering
    \includegraphics[width=0.85\linewidth]{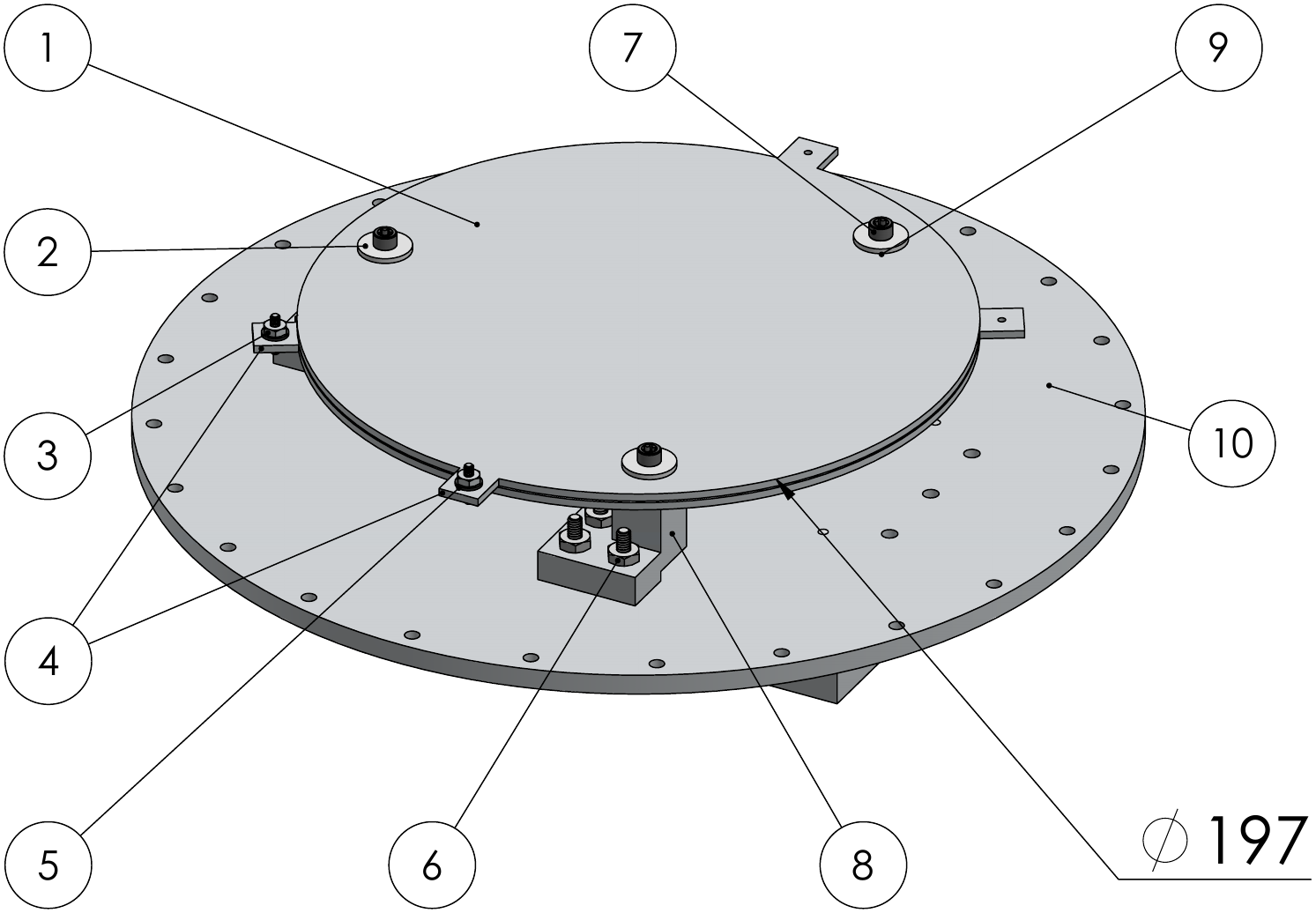}
    \caption{CAD drawing of the capacitor assembly. Two circular Al~1100 plates (1) of diameter \(197~\mathrm{mm}\) are separated by a \(0.6~\mathrm{mm}\) vacuum gap. Sapphire washers (2) provide electrical isolation between screws, plates, and stands. Al~6061 studs and nuts (3) connect the capacitor to the inductor via the electrical terminals on each plate (4), with tantalum washers (5) used to secure the NbTi wire from the inductor. For mounting, Al~6061 studs and nuts (6) are employed. The two plates are joined at three points \(120^{\circ}\) apart by tantalum screws (7), which also secure them to the Al~1100 base stands (8). Thin indium gaskets (9) are inserted at each washer-metal interface to improve thermal conductance. The completed assembly attaches to the chamber ceiling (10).
}
    \label{fig:capacitor_CAD}
\end{figure}
% \begin{figure}
%     \centering
%     \includegraphics[width=0.55\linewidth]{pics/CapacitorChamberTopAssemCircularUPSIDEDOWN_cross-section-cropped.pdf}
%     \caption{CAD cross section of the joint between two Al~1100 capacitor plates (3). The plates are coupled to each other and mounted to the supports (5) with a tantalum No.~8-32 UNC screw (1; shank diameter \(4.2\,\mathrm{mm}\)). Electrical isolation between the screw, plates, and supports is provided by three sapphire washers (2), each \(1.6\,\mathrm{mm}\) thick with a \(4.8\,\mathrm{mm}\) inner diameter (ID). The central washer is seated in opposing circular recesses (\(0.5\,\mathrm{mm}\) deep) machined into the facing plate surfaces for alignment and stability. Plate through-holes are intentionally oversized (\(6.8\,\mathrm{mm}\) diameter) relative to the screw to prevent electrical contact and accommodate thermal contraction. Thin indium gaskets are placed at each washer-plate interface to enhance thermal conductance.}
%     \label{fig:capacitor_cross-section_CAD}
% \end{figure}
% \begin{figure}
%     \centering
%     \includegraphics[width=0.75\linewidth]{pics/capacitor_photo.jpg}
%     \caption{Photograph of the capacitor mounted to the ceiling of the capacitor chamber. The assembly is shown upside-down for clarity.%
%     }
%     \label{fig:enter-label}
% \end{figure}
The capacitor is a circular parallel-plate device with a \(0.6\,\mathrm{mm}\) vacuum gap. Each plate (diameter \(197\,\mathrm{mm}\)) is machined from Al~1100, and the two plates are joined at three points \(120^\circ\) apart using tantalum screws with sapphire washers for electrical isolation (see Fig.~\ref{fig:capacitor_CAD}). An earlier design used Al~6061 plates, PTFE (Teflon) screws with alumina washers, and a square geometry.\cite{Kolevatov:Princeton:2024} As with the inductor frame, the present design was adopted because Al~1100 exhibits a sharper superconducting transition and higher \(T_c\) than Al~6061, and sapphire has a much lower loss tangent than PTFE and alumina. Replacing PTFE screws with tantalum hardware also provides a more robust, higher-force clamp.

To avoid electrical contact between the capacitor electrodes and to ground, the screw shanks are isolated from the plates and mounts, and a central sapphire washer is seated in shallow opposing counterbores on the facing plate surfaces. Additional sapphire washers—placed under the screw head and beneath the bottom plate—preserve electrical isolation while providing thermal conduction through the indium interfaces; see Fig.~3.11 in Ref.~\onlinecite{Kolevatov:Princeton:2024}.

\subsection{Magnetic Shielding}
\begin{figure}
    \centering
    \includegraphics[width=1\linewidth]{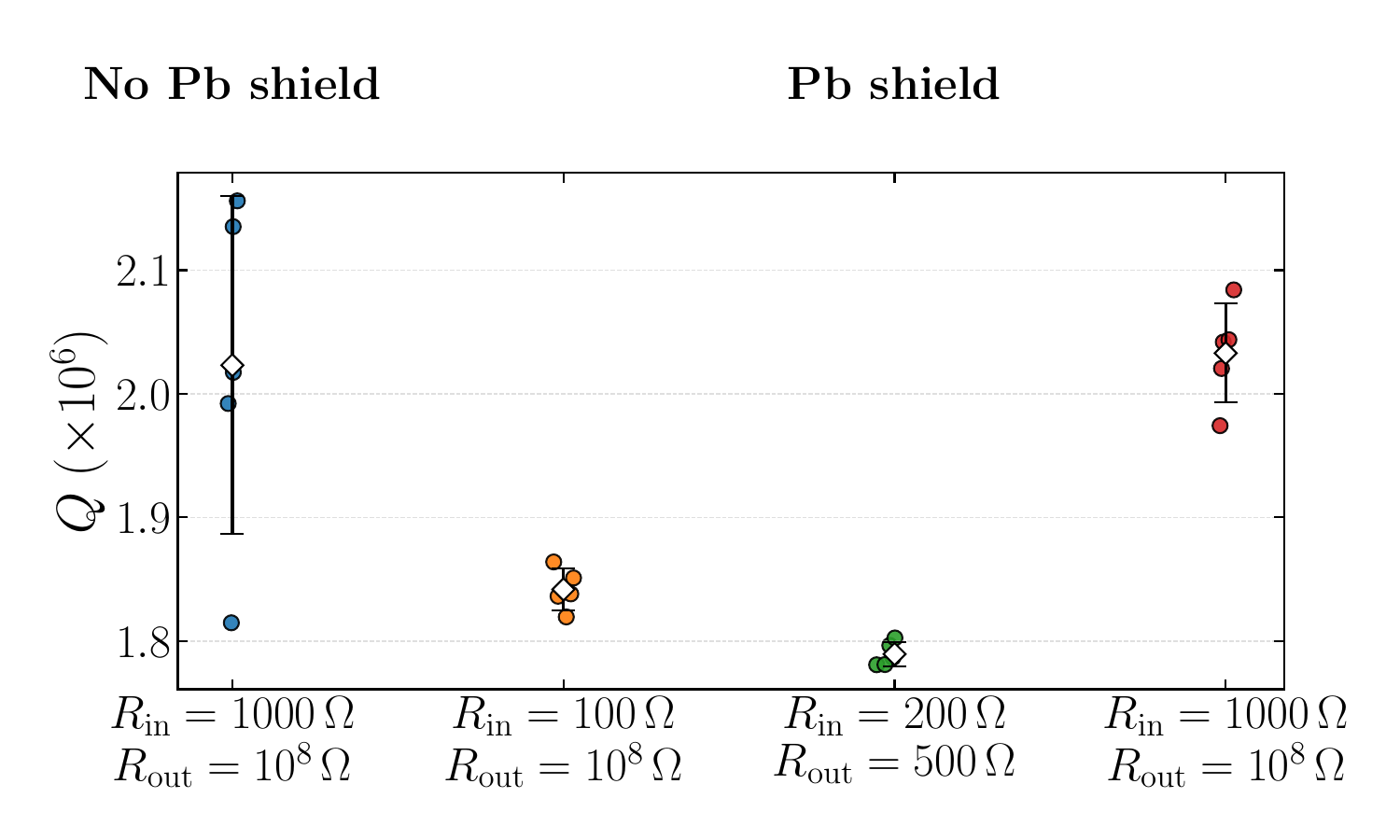}
    \caption{Quality factor \(Q\) for the resonator without lead shielding (blue) at \(R_\mathrm{in}=1~\mathrm{k}\Omega\), \(R_\mathrm{out}=100~\mathrm{M}\Omega\), and for the three input/output load configurations of Eq.~(\ref{eq:RinRout_configs}) with the lead magnetic shield (orange, green, red). For each configuration, five \({}^{3}\mathrm{He}/{}^{4}\mathrm{He}\) refrigerator cycles were taken, and the data were acquired at \(\approx 315\,\mathrm{mK}\); per-cycle values (colored dots) are obtained from coherent averages of demodulated, phase-aligned ringdown traces within that cycle (see Sec.~\ref{sec:meas_protocol}). White diamonds mark the mean across the five cycles; error bars denote the cycle-to-cycle standard deviation. The \(Q\) values measured without the lead shield exhibit larger cycle-to-cycle scatter, consistent with the trapped-flux hypothesis.
}
    \label{fig:Q_strip}
\end{figure}
Special attention is given to magnetic shielding. The cryostat provides five nested shells that contribute to magnetic attenuation in addition to their cryogenic functions. From outside to inside, these are: an outer vacuum shell (Al~6061), a \(300\,\mathrm{K}\) \(\mu\)-metal\cite{amunealMagneticShielding} shield, a \(40\,\mathrm{K}\) aluminum shield, a \(4\,\mathrm{K}\) Cryoperm\cite{amunealMagneticShielding} shield, and a \(4\,\mathrm{K}\) radiation shield that surrounds the high-purity aluminum resonator shield.

After observing variation in \(Q\) across sorption refrigerator cycles, the \(4\,\mathrm{K}\) radiation shield was wrapped with flexible lead sheets on the outside; seams were joined with Sn-Pb solder to form a continuous layer. This additional shielding improved the reproducibility of the measured quality factor \(Q\) between \(^3\)He/\(^4\)He refrigerator cycles. A plausible explanation is that the lead-clad \(4\,\mathrm{K}\) radiation shield \((T_c \approx 7.2\,\mathrm{K})\) maintains a superconducting enclosure throughout refrigerator recycling,\cite{Pearson1958} during which the resonator apparatus cycles between \(\approx 4\,\mathrm{K}\) and \(\approx 315\,\mathrm{mK}\). Because the Pb layer enters the Meissner state well above the transition temperatures of the resonator’s aluminum components, the resonator passes through the aluminum transitions \((T_c \approx 1.2\,\mathrm{K})\) while already screened from ambient \(\mathrm{dc}\) fields. In the earlier configuration (without Pb), it crossed the aluminum \(T_c\) in the presence of the Earth’s field \((\sim 25\text{-}65\,\mu\mathrm{T})\), leading to cycle-to-cycle variations in trapped flux and, consequently, in \(Q\). Similar \(Q\) degradation due to trapped magnetic flux has been reported in the SRF-cavity literature.\cite{Victor1968,Romanenko2014} The improved reproducibility after adding the Pb layer is consistent with this picture. Figure~\ref{fig:Q_strip} compares \(Q\) before and after the upgrade.

\subsection{Measurement Protocol}
\label{sec:meas_protocol}
\begin{figure}
    \centering
    \includegraphics[width=1\linewidth]{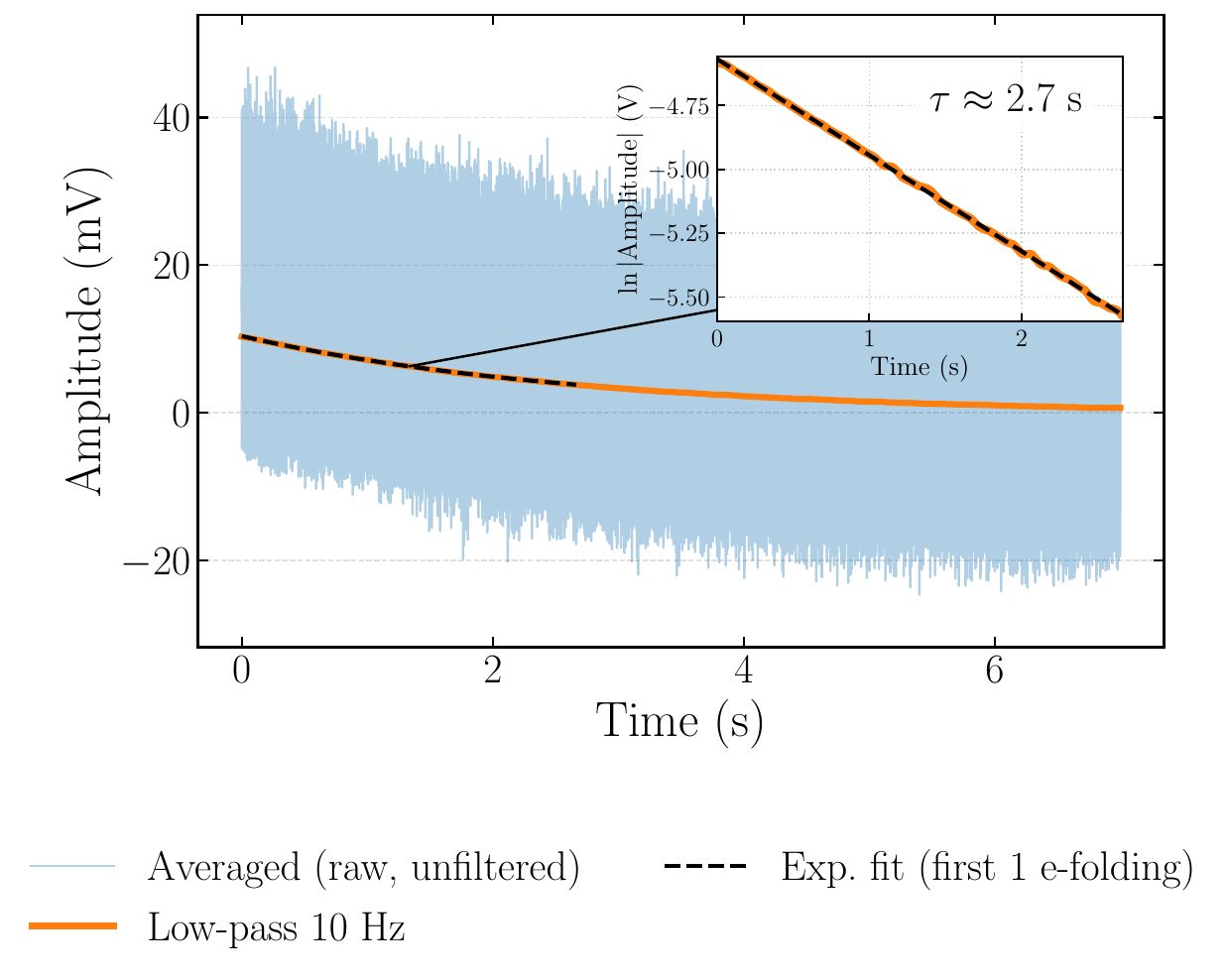}
    \caption{Ringdown amplitude envelope (blue) obtained by coherently averaging \(N_\mathrm{r}=29\) DC-demodulated, phase-aligned to zero ringdown traces (see Sec.~\ref{sec:meas_protocol}) for the \(R_\mathrm{in}=1~\mathrm{k}\Omega\), \(R_\mathrm{out}= 100~\mathrm{M}\Omega\) load configuration. The envelope is then filtered with a fifth-order Butterworth low-pass at \(10~\mathrm{Hz}\) (orange). The filtered envelope is log-transformed and its first e-folding is fitted with a straight line; the corresponding (back-transformed) exponential fit is shown. The inset shows the same filtered data in \(\ln\)-amplitude together with the linear fit over the fit window. The fit yields \(\tau\approx2.7~\mathrm{s}\) and \(f_{0}\approx249.7~\mathrm{kHz}\) (mean from Lorentzian fits to PSDs of the \(N_\mathrm{r}=29\) ringdowns), giving \(Q_{\mathrm{avg}}=\pi\, f_{0}\,\tau\approx 2.1\times 10^{6}\).}
    \label{fig:ringdown}
\end{figure}
The quality factor was measured using a ringdown (free-decay) method. The resonator was excited with a short sinusoidal burst detuned below resonance, \(f=f_{0}-\Delta f=249\,960~\mathrm{Hz}\), with \(\Delta f\approx 304~\mathrm{Hz}\).
 
After the drive is switched off, the circuit oscillates at \(f_{0}\) with an exponentially decaying envelope. The measured signal is modeled as
\begin{equation}
s(t) = V_{0}\, e^{-t/\tau} \cos\!\left(\omega_{0} t + \phi_{0}\right) + n(t),
\label{eq:ringdown}
\end{equation}
where \(V_{0}\) is the initial amplitude, \(\tau\) the decay constant, \(\omega_{0} = 2\pi f_{0}\) the angular resonance frequency, \(\phi_{0}\) the initial phase, and \(n(t)\) the noise. The quality factor is related to the time constant by
\begin{equation}
Q = \pi f_{0}\tau .
\label{eq:Q_ringdown}
\end{equation}

A typical setting uses an Agilent EDU33210A function generator to apply \(V_\mathrm{in}=1.2~\mathrm{V}\) single-ended through a \(1~\mathrm{k}\Omega\) series bias resistor (plus input-wire resistance \(R_\mathrm{wire}\approx 47~\Omega\)); the other end of the injection loop is shorted to chassis. This yields an injection-loop current \(I_\mathrm{in}\approx 1~\mathrm{mA}\) inductively coupled to the resonator via the injection loop (mutual inductance \(M_\mathrm{in}\); see Table~\ref{tab:R_params}). The voltage coupled into the resonator is
\(V_\mathrm{in,res}=\omega M_\mathrm{in} I_\mathrm{in}\approx 0.09~\mathrm{mV}\). The burst consists of \(N_\mathrm{cyc}=50{,}000\) sine cycles, with duration \(T_\mathrm{b}=N_\mathrm{cyc}/f\approx 0.20~\mathrm{s}\).

The output signal is read out differentially and amplified at room temperature by two Stanford Research Systems SR560 preamplifiers in series: the first stage (differential) provides a gain of \(10^{3}\), and the second stage (single-ended) a gain of \(20\), for a total gain of \(2\times10^{4}\). The SR560 presents a high input impedance (\(\sim 100\,\mathrm{M}\Omega\)), so the resonator output line is effectively open. Both stages are AC coupled and configured in internal band-pass mode (high-pass \(10\,\mathrm{kHz}\), low-pass \(300\,\mathrm{kHz}\)) with single-pole roll-offs (6 dB/octave, 20 dB/decade). In addition, a \(220\,\mathrm{kHz}\) high-pass filter (Thorlabs EF503) is inserted after the first SR560 and a \(240\,\mathrm{kHz}\) low-pass filter (Thorlabs EF504) after the second, defining a narrow \(220\text{-}240\,\mathrm{kHz}\) band-pass. The filtering prevents noise from saturating the digitizer.

The amplified signal is digitized with a Spectrum M2p.5966\text{-}x4 PCIe digitizer and processed on a host computer. The sampling rate is \(f_s=10\,\mathrm{MS/s}\) with a record length \(N_s=128\times1024\times1024=134{,}217{,}728\) samples, giving a total acquisition time \(T_\mathrm{rec}=N_s/f_s\approx 13.42\,\mathrm{s}\). A TTL marker from the function generator, issued at the end of the burst, provides the trigger. The digitizer runs in pre/post-trigger mode with \(25\%\) pre-trigger memory \((T_\mathrm{pre}\approx 3.36\,\mathrm{s})\) and \(75\%\) post-trigger memory \((T_\mathrm{post}\approx 10.06\,\mathrm{s})\), so the record includes baseline preceding the trigger and the subsequent ringdown.

The resonance frequency \(f_{0}\) is determined by fitting the power spectral density (PSD) of the measured signal \(s(t)\) to a Lorentzian,
\begin{equation}
\label{eq:lorentz_psd}
S(f)=C+\frac{A}{1+\bigl[2\pi\tau\,(f-f_{0})\bigr]^{2}},
\end{equation}
where \(A\) is the peak amplitude, and \(C\) is the flat noise background. A Lorentzian is appropriate because it describes the frequency response of a damped harmonic oscillator near resonance. The fit window spans eight expected linewidths (\(\Delta f_{\mathrm{FWHM}}\approx 0.125~\mathrm{Hz}\)) centered on the PSD peak, corresponding to \(|f-f_{0}|\le 4\,\Delta f_{\mathrm{FWHM}} = 0.5~\mathrm{Hz}\). The resonator parameters are defined in Table~\ref{tab:defs}.

The drive parameters (detuning \(\Delta f\equiv f_{0}-f\), injection amplitude \(V_{\mathrm{in}}\), and burst duration \(T_{\mathrm{b}}\)) are chosen to maximize SNR while avoiding self-heating. Detuning the drive and limiting \(V_{\mathrm{in}}\) reduce on-resonance power transfer and transient Joule heating; increasing \(V_{\mathrm{in}}\) or decreasing \(\Delta f\) improves SNR but risks overheating and excess ohmic losses that reduce \(Q\). Consequently, a small excitation is used, which increases the relative contribution of the noise \(n(t)\) in Eq.~(\ref{eq:ringdown}); to recover SNR, the ringdown traces are demodulated to DC, phase-aligned to zero, and coherently averaged, as detailed below.

Once the resonance frequency is obtained, a single-point discrete Fourier transform (DFT) is evaluated at \(f_{0}\),
\begin{equation}
S(f_{0}) = \sum_{n=0}^{N_{s}-1} s(n\Delta t)\, e^{-i 2\pi f_{0} n \Delta t},
\label{eq:single_dft}
\end{equation}
where \(\Delta t = 1/f_{s}\) is the sampling interval; the phase of the resonant mode is \(\phi_{0} = \arg\!\bigl(S(f_{0})\bigr)\). With \(f_{0}\) and \(\phi_{0}\) determined, the signal is demodulated to DC and phase-aligned to zero,
\begin{equation}
z(t) = s(t)\, e^{-i 2\pi f_{0} t}\, e^{-i \phi_{0}}
\label{eq:demod}
\end{equation}
yielding the complex demodulated trace \(z(t)\). Its real part contains the exponentially decaying envelope plus a residual noise term,
\begin{equation}
\Re\{z(t)\}=V_{0}\,e^{-t/\tau}+n'(t),
\label{eq:envelope}
\end{equation}
and coherent averaging over \(N_{\mathrm{r}}\) ringdowns suppresses the incoherent noise. To quantify the improvement, we compute a power SNR from the PSD. The signal power is obtained by integrating the PSD over a narrow band around the resonance: for the demodulated trace we integrate the PSD from \(0\) to \(2.5\,\Delta f_{\mathrm{FWHM}}\), while for each raw trace we integrate the PSD within \(|f-f_{0}|\le 2.5\,\Delta f_{\mathrm{FWHM}}\) about the resonance peak. The noise level is estimated as the median PSD in sidebands away from the peak, taken over frequency offsets from \(250\,\Delta f_{\mathrm{FWHM}}\) to \(2~\mathrm{kHz}\) from the resonance (using both sidebands for the raw traces). The corresponding noise power in the signal band is then taken as this median noise level multiplied by the signal bandwidth, and the SNR is defined as the excess signal power divided by this estimated noise power. Using this definition, coherent averaging of \(N_{\mathrm{r}}=29\) traces increases the SNR by a factor of \(24\): the final averaged, demodulated trace has \(\mathrm{SNR} = 57.4\)~dB, while the mean SNR of the individual raw traces is \(43.5\)~dB. The modest discrepancy relative to the ideal factor of 29 is plausibly due to imperfect coherence (e.g., small residual phase/frequency errors in the demodulation and phase-alignment procedure), correlated noise components, and excess low-frequency noise that partially contaminates the demodulated signal.

As a final step, the demodulated amplitude envelope (DC) is low-pass filtered with a fifth-order Butterworth filter with cutoff \(f_{c}=10~\mathrm{Hz}\), implemented in Python using \texttt{scipy.signal} (\texttt{butter} and \texttt{sosfiltfilt} for zero-phase forward-backward filtering).

The demodulated, low-pass-filtered envelope is log-transformed, and its first e-folding (to suppress noise-dominated tails) is fitted with a straight line; defining \(y(t)=\Re\!\{z(t)\}\propto e^{-t/\tau}\), the slope of \(\ln y(t)\) is
\begin{equation}
m=-\frac{1}{\tau},
\label{eq:tauslope}
\end{equation}
so that \(\tau=-1/m\). The quality factor then follows from Eq.~(\ref{eq:Q_ringdown}), with \(f_{0}\) taken as the mean resonance frequency across all \(N_{\mathrm{r}}\) ringdowns.

An illustration of the amplitude envelope obtained by this method, from one of the \({}^{3}\mathrm{He}/{}^{4}\mathrm{He}\) cycles with \(R_{\mathrm{in}}=1~\mathrm{k}\Omega\) and \(R_{\mathrm{out}}= 100~\mathrm{M}\Omega\) (see Sec.~\ref{sec:analysis_res}), together with the resulting fit, is shown in Fig.~\ref{fig:ringdown}.

As an additional consistency check, the PSDs of all individual ringdowns are also averaged and fitted with the Lorentzian in Eq.~(\ref{eq:lorentz_psd}), yielding \(Q_{\mathrm{PSD}}=\pi f_{0}\tau\), with \(\tau\) obtained from the fit. This PSD-based estimate is intrinsically less precise because the PSD frequency resolution is \(\delta f=f_{s}/(0.75\,N_{s})\approx 0.099~\mathrm{Hz}\) (75\% post-trigger), i.e., roughly one point per expected linewidth (\(\Delta f_{FWHM}\approx 0.125~\mathrm{Hz}\)); accordingly, it is used only as a cross-check, while the time-domain approach is the primary method. Even so, the PSD-derived \(Q\) agrees with the time-domain value to within \(\lesssim 5\%\).

\section{Analysis and results}
\label{sec:analysis_res}
\begin{figure}
    \centering
    \includegraphics[width=1\linewidth]{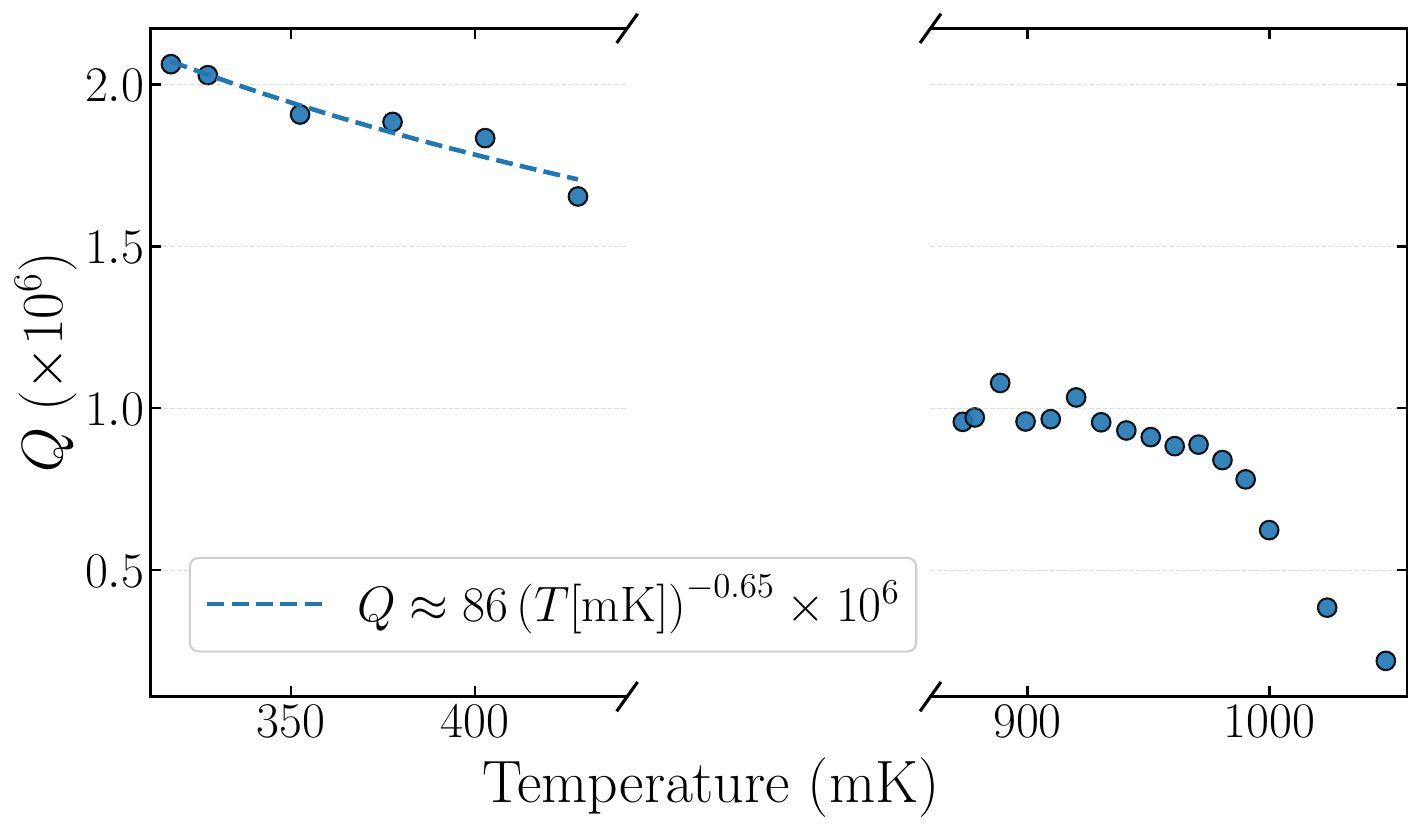}
    \caption{Quality factor \(Q\) versus temperature. At each temperature, \(N_\mathrm{r}=5\) consecutive ringdowns were acquired for the \(R_\mathrm{in}=1~\mathrm{k}\Omega\), \(R_\mathrm{out}=100~\mathrm{M}\Omega\) load configuration; the plotted points (dots) are obtained by coherent averaging of the DC-demodulated, zero-phase-aligned traces (see Sec.~\ref{sec:meas_protocol}) at the \(\approx10~\mathrm{mK}\) temperature steps (\(\approx25~\mathrm{mK}\) for \(T<425~\mathrm{mK}\)). After stepping the copper plate temperature to a new setpoint, the system was allowed to re-equilibrate for \(90~\mathrm{min}\) (\(30~\mathrm{min}\) for \(T<425~\mathrm{mK}\)) before new acquisition. The large decrease in \(Q\) above 1 Kelvin is due to approaching the superconductor-to-normal metal transition in aluminum.}
\label{fig:Q_vs_temperature}
\end{figure}
A set of runs was performed at base temperature of \({}^{3}\mathrm{He}/{}^{4}\mathrm{He}\) cycle (\(\approx 315\,\mathrm{mK}\)) to extract the unloaded resistance \(R_\mathrm{ul}\) and the unloaded quality factor \(Q_\mathrm{ul}\) at the coldest resonator temperature.

Ringdowns were taken under several known input/output loads to obtain the loaded \(Q\) and the corresponding loaded (residual) resistance \(R=\omega_{0}L/Q\). The three load configurations (indexed by \(j=1,2,3\)) were
\begin{equation}
\begin{array}{ll}
(j{=}1): & R_{\mathrm{in}}=100~\Omega,\quad R_{\mathrm{out}}= 100~\mathrm{M}\Omega,\\
(j{=}2): & R_{\mathrm{in}}=1~\mathrm{k}\Omega,\quad R_{\mathrm{out}}= 100~\mathrm{M}\Omega,\\
(j{=}3): & R_{\mathrm{in}}=200~\Omega,\quad R_{\mathrm{out}}=500~\Omega.
\end{array}
\label{eq:RinRout_configs}
\end{equation}
Here \(R_{\mathrm{out}}= 100~\mathrm{M}\Omega\) is set by the SR560 input impedance, and \(R_{\mathrm{out}}=500~\Omega\) is implemented by shunting \(500~\Omega\) across the SR560 input.

For each configuration \(j\), five independent \({}^{3}\mathrm{He}/{}^{4}\mathrm{He}\) refrigerator cycles (\(k=1,\ldots,5\)) were performed, for a total of 15 cycles. After stabilizing at the base-temperature plateau (\(\approx 315~\mathrm{mK}\)), ringdowns were acquired over \(\approx 3~\mathrm{h}\) at \(\approx 6~\mathrm{min}\) intervals to allow re-equilibration, yielding \(N_\mathrm{r}=29\) traces per cycle.

For each cycle \(k\) within configuration \(j\), the post-processing procedure of Sec.~\ref{sec:meas_protocol} was applied. All ringdown traces acquired in that cycle were phase-aligned to zero and demodulated to DC; traces were then coherently averaged, and the resulting envelope was log-transformed and fit with a straight line to obtain the slope \(m_{jk}\equiv -1/\tau_{jk}\). The quality factor and the residual (loaded) resistance then follow from that slope:
\begin{equation}
Q_{jk}=-\frac{\pi f_{0,jk}}{m_{jk}}, \qquad R_{jk}=-\,2L\,m_{jk},
\label{eq:QR_from_m}
\end{equation}
where \(L\pm \sigma_{L}=(750\pm1)\,\mu\mathrm{H}\) (from prior calibration) and \(f_{0,jk}\) is the per-cycle mean resonance frequency obtained from Lorentzian fits to the power spectral densities (PSDs) of the individual traces in that cycle. (Since \(m_{jk}<0\) for a decaying envelope, Eq.~(\ref{eq:QR_from_m}) yields \(Q_{jk}>0\) and \(R_{jk}>0\).)

Results for \(Q_{jk}\) for each cycle, together with the cycle-to-cycle standard deviation within each configuration defined in Eq.~(\ref{eq:RinRout_configs}), are shown in Fig.~\ref{fig:Q_strip}. For reference, the figure also includes data from a previous cooldown performed before installation of the lead magnetic shield. 

%For each configuration \(j\) defined in Eq.~(\ref{eq:RinRout_configs}), the residual (loaded) resistance was taken as the (unweighted) mean of the per-cycle values:
%\[
%R_{j}=\frac{1}{K}\sum_{k=1}^{K} R_{jk}, \qquad K=5.
%\]
%Its uncertainty \(\sigma_{R_{j}}\) is the quadrature sum of the standard error from the observed cycle-to-cycle scatter, \(s_{j}/\sqrt{K}\), and the standard error from the individual per-cycle uncertainties, \((1/K)\sqrt{\sum_{k=1}^{K}\sigma_{R_{jk}}^{2}}\); here \(s_{j}\) is the standard deviation of \(\{R_{jk}\}\) for configuration \(j\), and \(\sigma_{R_{jk}}\) includes fit contributions via the slope-fit uncertainty \(\sigma_{m,jk}\) and the inductance calibration \(\sigma_{L}\).

%Since the resonance frequency is configuration-independent, the final value was taken as the mean over all cycles:
%\[
%\bar f_{0}=\frac{1}{N}\sum_{j=1}^{J}\sum_{k=1}^{K} f_{0,jk}, 
%\qquad J=3,\; K=5,\; N=JK=15.
%\]
%Its statistical uncertainty \(\sigma_{\bar f_{0}}\) is the quadrature sum of the standard error from the overall scatter, \(s_{f}/\sqrt{N}\), and the standard error from the per-cycle frequency uncertainties, \((1/N)\sqrt{\sum_{j=1}^{J}\sum_{k=1}^{K}\sigma_{f_{0},jk}^{2}}\) (assuming the errors from each cycle are uncorrelated); here \(s_{f}\) is the standard deviation of \(\{f_{0,jk}\}\) over all \(N\) cycles, and \(\sigma_{f_{0},jk}\), is the per-cycle frequency uncertainty (from the Lorentzian-fit dispersion within that cycle).

The residual (loaded) resistance and resonance frequency were evaluated by computing the mean and standard deviation of per-cycle values. Using the three loaded resistances \(R_{j}\) with uncertainties \(\sigma_{R_{j}}\) for \(j=1,2,3\) (configurations in Eq.~(\ref{eq:RinRout_configs})) and the common resonance frequency \(\bar f_{0}\) with uncertainty \(\sigma_{\bar f_{0}}\), Eq.~(\ref{eq:R_resonator}) was fitted by linear least squares to obtain the unloaded resistance \(R_{\mathrm{ul}}\). The uncertainty \(\sigma_{R_{\mathrm{ul}}}\) was taken as the square root of the corresponding diagonal element of the parameter-covariance matrix. Finally, the unloaded quality factor was computed as
\begin{equation}
\label{eq:Qul_def}
Q_{\mathrm{ul}}=\frac{\omega L}{R_{\mathrm{ul}}}, \qquad \omega=2\pi\bar f_{0},
\end{equation}
and its uncertainty was obtained by linear error propagation of \(\sigma_{\bar f_{0}},\,\sigma_{L}\), and \(\sigma_{R_{\mathrm{ul}}}\).

The unloaded resistance, representing internal resonator loss, was measured to be \(R_{\mathrm{ul}}=0.572\pm0.005~\mathrm{m}\Omega\). This corresponds to
\begin{equation}
\label{eq:Q_unloaded}
Q_{\mathrm{ul}}=(2.06\pm0.02)\times10^{6},
\end{equation}
at \(T\approx315~\mathrm{mK}\), with \(\bar f_{0}\approx 249\,656~\mathrm{Hz}\).

Additionally, the quality factor \(Q\) was measured as a function of temperature for a single load configuration, \(R_\mathrm{in}=1\,\mathrm{k}\Omega\) and \(R_\mathrm{out}\approx 100\,\mathrm{M}\Omega\). The resonator was first cooled to the base temperature of the \({}^{3}\mathrm{He}/{}^{4}\mathrm{He}\) cycle (\(\approx 315\,\mathrm{mK}\)), then warmed in a controlled manner by applying heat to the copper resonator plate under closed-loop PID control using a Lake Shore Model 372 AC Resistance Bridge. The results are shown in Fig.~\ref{fig:Q_vs_temperature}.

A trend of \(Q\) increasing with decreasing temperature is observed, despite operation well below the superconducting transition temperatures of all superconducting components. This behavior is unlikely to be explained by two-level-system (TLS) dielectric loss, since standard TLS loss models typically predict increasing (or saturating) dissipation as the temperature is reduced in the relevant regime. \cite{frossati1977spectrum,GaoThesis2008,Pappas2011TLS,Zmuidzinas2012Microresonators}
A more plausible interpretation is that the dissipation is dominated by conductor loss consistent with BCS/Mattis--Bardeen electrodynamics, for which the surface resistance decreases as the temperature is reduced.\cite{MattisBardeen1958} Another possibility is incomplete thermalization of one or more resonator components, such that an effective device temperature differs from the bath temperature.

Residual trapped magnetic flux can also contribute to dissipation and may introduce temperature-dependent losses through flux dynamics and pinning. In particular, trapped flux can respond to the oscillating currents and fields of the resonator producing an additional effective surface resistance. \cite{Victor1968,Romanenko2014,PhysRevLett.67.386} The magnitude of this contribution depends on the flux-flow viscosity and pinning strength, both of which can vary with temperature. \cite{park1992vortex}

A quantitative model may also need to include modifications of the transition temperature and quasiparticle density of states due to impurities in aluminum alloys \cite{chanin1959impurity,o2010quasiparticle} and independent measurements of the effective temperature relevant to dissipation in the resonator. A quantitative explanation of this \(Q(T)\) behavior is beyond the scope of the present work.

\section{Conclusion}
A superconducting lumped-element resonator of $\sim 1$ liter volume with fixed resonance frequency \(f_{0}\approx 250~\mathrm{kHz}\) and, to our knowledge, an unprecedented unloaded quality factor \(Q_\mathrm{ul}\approx 2.1\times10^{6}\) has been demonstrated. Development of the apparatus yielded the following practical guidelines:

\begin{itemize}
\item \textbf{Dielectrics.} High \(Q\) is obtained only with ultralow-loss dielectrics; sapphire should be used whenever practical, while higher-loss alternatives such as alumina should be avoided.
\item \textbf{Conductors.} Choose the highest-purity, highest-\(T_c\) superconducting metals available. Between alloys of the same base metal, the purer grade is preferred (e.g., Al~1100 outperformed Al~6061).
\item \textbf{Coupling.} External loading must be limited while maintaining sufficient drive and readout. In this implementation, mutual inductive couplings of \(M_{\mathrm{in}}\approx 52~\mathrm{nH}\) and \(M_{\mathrm{out}}\approx 103~\mathrm{nH}\) provided an effective compromise between minimal loading and adequate signal-to-noise.
\item \textbf{Electrical joints.} Joints can dominate loss. Minimize their number (e.g., connect the inductor directly to the capacitor, avoiding intermediate terminals) and use superconducting screw-terminals for S-S contacts. Prepare contact surfaces prior to assembly (for Al, light abrasion with SiC paper followed by solvent cleaning: acetone, then isopropanol, and dry).
\item \textbf{Fasteners/torque.} Use consistent, calibrated torque settings for all fasteners to ensure reproducibility, good thermalization, and reliable electrical contact at joints.
\item \textbf{Magnetic environment.} Magnetic shielding is crucial to mitigate trapped flux. If any component may cross \(T_c\) during operation, an outer superconducting shield (lead, in this implementation) that remains superconducting can reduce flux trapping and trial scatter in measured \(Q\).
\item \textbf{Measurement procedure.} The \(Q\)-measurement is sensitive to drive conditions. The injection current should be adjusted, and the drive frequency slightly detuned from resonance to obtain a resolvable signal while avoiding Joule heating that transiently increases losses and biases \(Q\) low. Operation in the linear, non-heating regime is essential for reproducibility.
\item \textbf{Temperature.} Even well below \(T_c\) for all components, the measured \(Q\) showed temperature dependence, indicating that the lowest attainable temperature benefits \(Q\)-performance.
\end{itemize}
Several avenues can extend this work. First, pursue lower temperatures: because \(Q\) increases as \(T\) decreases even well below all components' \(T_c\), operating in a dilution refrigerator to reach the \(\mathrm{mK}\) regime should further improve performance. Second, reduce dielectric loss by eliminating Formvar insulation on the NbTi wire (\(\tan\delta\sim10^{-3}\)), for example, by using an uninsulated winding.\cite{Ulmer2009} Third, upgrade materials by replacing Al~1100 components with niobium to exploit its higher critical temperature \(T_c\approx9.3~\mathrm{K}\)\cite{Webb2015} and to improve thermal-contraction matching to tantalum hardware; for a cooldown from \(\sim300\) to \(0~\mathrm{K}\), the integrated linear contractions are \(\Delta L/L\approx4.15\times10^{-3}\) for Al versus \(\approx1.43\times10^{-3}\) for both Nb and Ta,\cite{NBSMono29} favoring Nb-Ta joints. Fourth, integrate the resonator output with a SQUID readout chain to achieve lower noise; the present apparatus already includes a first-stage SQUID in the lower chamber and a second-stage amplifier at \(4~\mathrm{K}\), facilitating system-level integration. SQUID readout resolves the thermal noise of the resonator, which allows the measurement of its effective temperature. Knowledge of this temperature, which may differ from the measured copper plate temperature (x-axis in Fig. \ref{fig:Q_vs_temperature}), is necessary to accurately model the physical origins of the resonator loss. Finally, implement a tunable capacitor for resonance-frequency control.
%; related design work is ongoing within the DMRadio collaboration.

The lessons learned in this R\&D can be directly incorporated into the design of low-mass axion searches.
%, including the DMRadio campaign. 

% If in two-column mode, this environment will change to single-column format so that long equations can be displayed. 
% Use only when necessary.
%\begin{widetext}
%$$\mbox{put long equation here}$$
%\end{widetext}

% Figures should be put into the text as floats. 
% Use the graphics or graphicx packages (distributed with LaTeX2e).
% See the LaTeX Graphics Companion by Michel Goosens, Sebastian Rahtz, and Frank Mittelbach for examples. 
%
% Here is an example of the general form of a figure:
% Fill in the caption in the braces of the \caption{} command. 
% Put the label that you will use with \ref{} command in the braces of the \label{} command.
%
% \begin{figure}
% \includegraphics{}%
% \caption{\label{}}%
% \end{figure}

% Tables may be be put in the text as floats.
% Here is an example of the general form of a table:
% Fill in the caption in the braces of the \caption{} command. Put the label
% that you will use with \ref{} command in the braces of the \label{} command.
% Insert the column specifiers (l, r, c, d, etc.) in the empty braces of the
% \begin{tabular}{} command.
%
% \begin{table}
% \caption{\label{} }
% \begin{tabular}{}
% \end{tabular}
% \end{table}

% If you have acknowledgments, this puts in the proper section head.
\begin{acknowledgments}
Development of the resonator took place over more than five years and involved many people. We thank Joelle-Marie Begin, Nate Otto, and Joseph Wiedemann for assistance in cryostat assembly and helpful discussions. We thank undergraduates Deniz Erdag, Nicky He, Rebecka Mahring, Ryan Marin, Paolo Montoya, Nathaniel Bruss, Oyu Enkhbold, Vivian Huang, Nastassia Patnaik, Pranav Vadapalli, Nicholas Callan, and Jessica Fox for assistance in various tasks during summer sessions. We thank graduate students Maksim Borovkov, Nicolas Patino, and Dmitrii Trunin for their work on experimental projects that informed the apparatus design. S.C. and R.K. thank Steve Lowe, Pat Bradshaw, Matt Komor, and Glenn Atkinson for invaluable support in the machine shop. We thank Kent Irwin for a careful reading of the manuscript. We thank members of the DMRadio collaboration for useful discussions. We thank the Editor and the Reviewers for their thoughtful suggestions, which improved the manuscript. This material was based upon work supported by the U.S. Department of Energy, Office of Science, Office of High Energy Physics, under Award Number DE-SC0007968. This work has been supported by Simons Foundation grant number MP-TMPS-00002996 and the ``Table-top experiments for fundamental physics'' program, sponsored by the Gordon and Betty Moore Foundation, Simons Foundation, Alfred P. Sloan Foundation, and John Templeton Foundation.  We gratefully acknowledge support from the Princeton Innovation Fund for New Ideas in the Natural Sciences. S.C. was supported by the R.H. Dicke Postdoctoral Fellowship. 
\end{acknowledgments}

\section*{Data Availability}
The data that support the findings of this study are available from the corresponding author upon reasonable request.

% Create the reference section using BibTeX:
\bibliographystyle{aipnum4-1}
\bibliography{literature.bib}

\end{document}